\documentclass{appolb}
\usepackage{graphicx}

\begin{document}
\title{A new event display for the KLOE-2 experiment%
}
\author{Wojciech Krzemie\'n \\ on behalf of the KLOE-2 Collaboration
\address{High Energy Physics Division, National Centre for Nuclear Research, \\
  05-400 Otwock-\'Swierk, POLAND.
}
}
\maketitle
\begin{abstract}
  In this contribution we describe KNEDLE - the new event 
  display for the KLOE-2 experiment. 
  The basic objectives and software requirements are presented. 
  The current status of the development 
  is given along with a short discussion of the future plans.

\end{abstract}
  
\section{Introduction}
The KLOE-2 is an upgraded version of the KLOE detector which is installed at DA$\Phi$NE, 
the e$^+$e$^-$ collider located at the Frascati Laboratories of INFN.
The newly installed subdetectors are (i) the Inner Tracker~\cite{inner0, inner} for 
the improvement of the vertex position resolution and the acceptance increase 
for low transverse momentum tracks; (ii) two pairs of small angle tagging 
devices for detection of low (Low Energy Tagger - LET~\cite{let}) and high (High Energy Tagger - HET~\cite{het}), 
energy electrons and positrons from $e^+e^- \rightarrow e^+e^-X$ reactions; (iii)
crystal calorimeters (CCALT) for covering of the low polar
angle region to increase acceptance for very forward
electrons and photons down to  8 deg~\cite{ccalt}, and a tile
calorimeter (QCALT) for detection of photons
coming from $K_L$ decays in the drift chamber~\cite{qcalt}. 
Currently, the intensive work involving commissioning of new detectors 
is being performed. 
At the same time, new programming procedures are being developed, 
e.g. reconstruction procedures for the Inner Tracker.

One of the tools that can be particularly useful  to quickly 
verify the correctness of reconstruction algorithms or to 
reveal some malfunctioning of the detectors parts
is the so-called event display. 
This program permits to graphically visualize 
reconstructed trajectories or energies of detected particles  
on event by event basis. 
In this contribution we present KNEDLE (\emph{Kloe New Event DispLay Environment}), the new event display for the KLOE-2 detector. 

\section{Comparison with DIDONE event display}
DIDONE (\emph{DAFNE Interactions Display}) 
is a previous event display for the KLOE detector. 
Its graphical layout is based on the OPACS graphical libraries written entirely in C language~\footnote{OPACS graphical libraries set was developed in the 90s in LAL Orsay. It was implemented using ANSI C and X11. This graphical package was adopted in many event displays like NA48, NEMO and KLOE experiments. }. 
It interfaces the content of several YBOS banks~\footnote{YBOS banks are the special data format used in KLOE data flow scheme. } 
created by the reconstruction program with the KLOE database. 
Although, DIDONE is still operational, e.g. 
particle trajectories reconstructed in the drift chamber can be visualized,
there are important arguments that favored the development of an entirely 
new display. 
The first argument is that the program is outdated. As it was mentioned before, 
its core is based on OPACS package system, 
which is an abandoned project, 
without any support available. 
The use of callback functions combined with a lack of documentation 
make the current code very hard to maintain.  
In addition, the DIDONE display runs only on AIX server, 
and a significant amount of time would be needed to make it compatible 
with modern Linux systems.
Therefore, the decision was made to stop the previous project, 
and to concentrate on the development of a new solution 
based on the ROOT libraries.

\section{Requirements}

The KNEDLE event display aims to replace and enhance the functionality of DIDONE. 
The application should contain the graphical user interface with the visualization 
of different detector components. 
It was decided  that in the first development stage, 
the event display for Inner Tracker, 
following the Drift Chamber part should be implemented,  
but the final version will include also other detector's parts.
To make it possible to run on personal computers and on the KLOE server, 
the program must work both on AIX and Linux operating systems.
Finally the appropriate documentation should be provided along 
the development of the project.

\section{Programming details}

The KNEDLE application is written in C++ with the object-oriented approach. 
It uses a several ROOT~\cite{root} libraries like Geometry library to implement detectors geometry. 
The graphical user interface is implemented using Qt-like GUI library.  
For the backward compatibility with the ROOT version available on the KLOE server
(ROOT 5.08 compiled with xlc++), 
some of the packages features were deliberately ommited. 
The application contains a configurable logger, 
that delivers the login information that can be printed on the screen
or saved to a special file. 
The quality of the code is assured by a set of unit tests.
The BOOST unit test environment~\cite{boost} was chosen for this purpose.
The code documentation can be generated using Doxygen~\cite{doxygen}. 
Finally, the whole source code is store on a git repository.

\subsection{Internal architecture}

\begin{figure}[htb]
\centerline{%
\includegraphics[width=10cm]{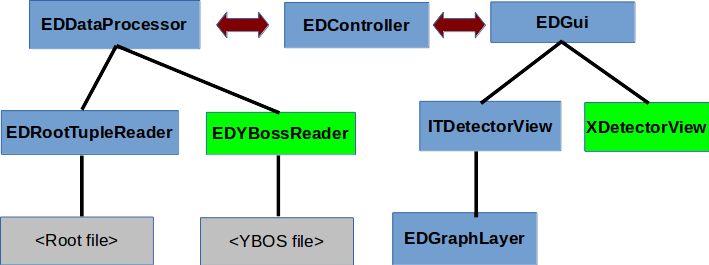}}
\caption{Simplified architecture of the KNEDLE application; only the most important modules are shown.
The class marked as blue rectangles are currently implemented, whereas the green rectangles represent
example extensions of the program. The grey rectangles denote input file formats.}
\label{Fig:archi}
\end{figure}

The KNEDLE internal architecture was designed in a modular form, in order to 
easily extend it with further detector components.
In order to separate a given graphical representation from data, 
the Model-View-Controller pattern has been applied (see Fig.~\ref{Fig:archi}).

The DataProcessor is responsible for reading the data and providing it 
for the visualization. The EDGui manges graphical user interface
along with the different detector views.
The Controller mediates between 
the DataProcessor and the EDGui module. The communication between the Controller
module and the user interface is implemented using the signal-slot technique. 
The EDRootTupleReader is responsible for reading the data in the tree ROOT format.

\section{Graphical Interface}

\begin{figure}[htb]
\centerline{%
\includegraphics[width=8cm]{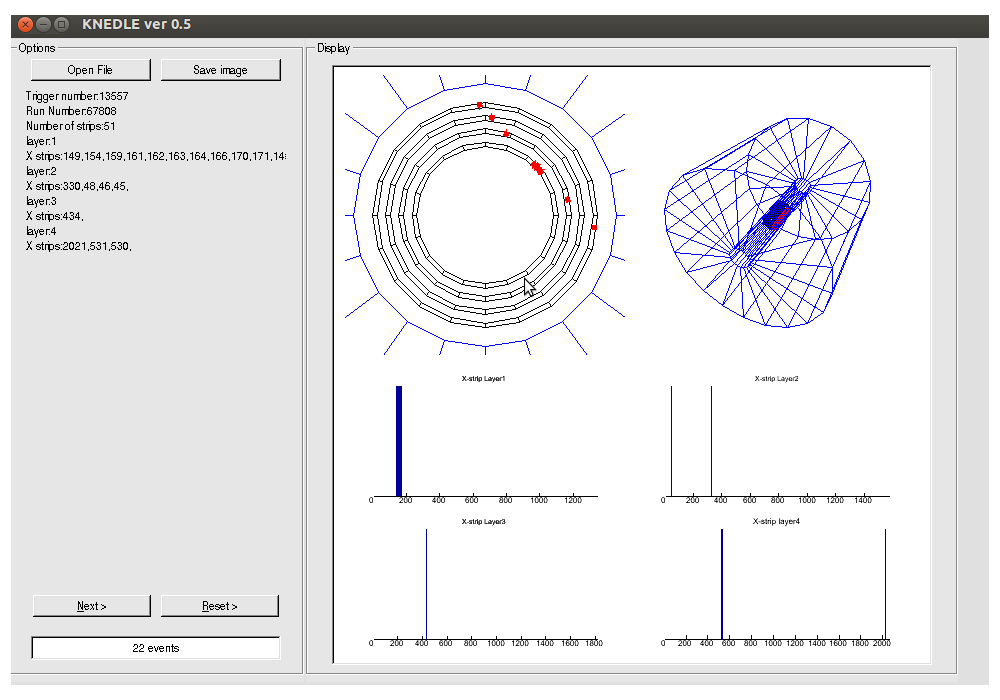}}
\caption{The graphical user interface of the KNEDLE. 
The panels on the right represent the view of the Inner Tracker layers and the Drift Chamber geometry.
The red dots correspond to the activated strips in the event. 
The bottom plots display the numbers of activated strips in layers. 
The panel on the left side contains information about the current event.}
\label{Fig:view}
\end{figure}

The current version of the event display implements the Inner Tracker
and Drift Chamber geometries. It permits for the visualization of the 
X-strips layers (for the geometry details please refer to~\cite{inner})
on the event by event basis. Along with the graphical representation, the basic
text information is displayed (see Fig.~\ref{Fig:view}). 

The button "Save to picture" permits to save the current view in the "png" format.

\section{Summary and Outlook}

In this article, we presented  KNEDLE, 
the new event display for the KLOE-2 detector.
It is a ROOT-based application written in C++,
that can be compiled and run on the KLOE server (AIX) and 
separately on the Linux operating system.
The current version implements several views of the Inner
Tracker and the Drift Chamber. It permits to visualize the Inner Tracker strips by layers.
Also the procedure to visualize the trajectories in the
Drift Chamber is ready to use.
The program supports the ROOT tree format as an input data.
The operation with the YBOS files is planned to be added in the near future.

\section{Acknowledgements}
This work was supported in part by the EU Integrated Infrastructure Initiative Hadron Physics Project under contract number RII3-CT- 2004-506078; by the European Commission under the 7th Framework Programme through the \textit{Research Infrastructures} action of the \textit{Capacities} Programme, Call: FP7-INFRASTRUCTURES-2008-1, Grant Agreement No. 227431; by the Polish National Science Centre through the Grants No. 0469/B/H03/ 2009/37, 0309/B/H03/2011/40, 2011/03/N/ST2/02641, 2011/01/D/ST2/ 00748, 2011/03/N/ST2/02652, 2013/08/M/ST2/00323, 2014/12/S/ST2/00459 and by the Foundation for Polish Science through the MPD programme and the project HOMING PLUS BIS/2011-4/3.
%
%
%


\begin{thebibliography}{00}
  \bibitem{inner0} A.~Balla et al.,{\it Acta Phys. Pol. B Proc. Suppl.}, Vol. 6 , No. 4, 1053 (2013).
  \bibitem{inner} A.~Di Cicco and G.~Morello, {\it Acta. Phys. Pol. B} this number (2014).
  \bibitem{let} D.~Babusci et al., {\it Nucl. Instr. \& Meth.} {\bf A617}, 81 (2010).
  \bibitem{het} F.~Archilli et al., {\it Nucl. Instr. \& Meth.} {\bf A617}, 266 (2010).
  \bibitem{ccalt} F.~Happacher et al., {\it Nucl. Phys. Proc. Suppl. 197}, 215 (2009).
  \bibitem{qcalt} M.~Cordelli et al., {\it Nucl. Instr. \& Meth.} {\bf A617}, 105 (2010).
  \bibitem{boost} BOOST - website of the project {http://www.boost.org/}
  \bibitem{doxygen} Doxygen - website of the project {www.doxygen.org}
  \bibitem{root} R.~Brun and F.~Rademakers, {\it Nucl. Inst. \& Meth. in Phys. Res.} {\bf A389} 81-86 (1997). See also {http://root.cern.ch/}.

\end{thebibliography}
\end{document}